\documentclass[aps,preprint]{revtex4}

\usepackage{graphicx}

\newcommand{\ch}{{\cal H}}
\newcommand{\ce}{{\cal E}}

\begin{document}

\date{\today}

\title{Gain and loss  in open quantum systems} 

\author{
Hichem Eleuch$^{1}$\footnote{email: hichemeleuch@tamu.edu} and 
Ingrid Rotter$^{2}$\footnote{email: rotter@pks.mpg.de, corresponding author}}

\address{
$^1$ Institute for Quantum Science and Engineering,
Texas A$\&$M University, College Station, Texas 77843, USA}
\address{
$^2$ Max Planck Institute for the Physics of Complex Systems,
D-01187 Dresden, Germany  }

\vspace*{1.5cm}

\begin{abstract}

Photosynthesis is the basic process used by plants 
to convert light energy  in reaction centers  
into chemical energy. The high efficiency of this process is 
not yet understood today. 
Using the formalism for the description of open quantum systems
by means of a non-Hermitian Hamilton operator, 
we consider initially the interplay of gain (acceptor) and 
loss (donor). Near singular points
it causes fluctuations of the cross section
which appear  without any excitation of  internal degrees of 
freedom of the system. This process
occurs therefore very quickly and with high efficiency. We then
consider the excitation
of  resonance states of the system by means of these fluctuations. 
This second step of the whole
process takes place much slower than the first one, because 
it involves the excitation of
internal degrees of freedom of the system. 
The two-step process as a whole is highly efficient and the decay
is bi-exponential. We provide, if possible, the results of 
analytical studies, otherwise  characteristic numerical results.
The similarities of the obtained results to light harvesting in 
photosynthetic organisms are discussed.

\end{abstract}

\maketitle

\section{Introduction}
\label{intr}

Photosynthetic organisms capture visible light 
in their light-harvesting complex and transfer
the excitation energy to the reaction center which stores 
the energy from the photon in 
chemical bonds. This process occurs with nearly perfect 
efficiency. The primary process 
occurring in the light-harvesting complex, 
is the exciton transfer between acceptor and donor,
while the transfer of the energy to the reaction center appears as  
a secondary process.
Both processes are nothing but two parts of the total 
light harvesting. 

A few years ago, evidence of coherent quantum energy transfer has been
found experimentally \cite{engel,engel2}.
Recent experimental results \cite{engel-ed}
demonstrated that photosynthetic bio-complexes exhibit collective
quantum coherence during primary exciton transfer processes 
that occur on the time scale of some hundreds of femtoseconds.
Furthermore, the coherence in such a system exhibits
a bi-exponential decay consisting of a slow component with a 
lifetime of hundreds of
femtoseconds and a rapid component with a lifetime of tens of
femtoseconds \cite{fleming}. The long-lived components are 
correlated with intramolecular  modes within the
reaction center, as shown experimentally  \cite{romero}.

These results  induced different theoretical considerations which are
related to the role of quantum coherence in 
the photosynthesis. For example, the equivalence of quantum and
classical coherence in electronic energy transfer is considered in
\cite{briggs}. In \cite{huelga}, the fundamental role of
noise-assisted transport
is investigated. In \cite{scully1}, it is shown that the efficiency
is increased by reducing radiative recombination due to quantum
coherence. The Hamiltonian of the system in these (and many other) 
papers is assumed to be
Hermitian although photosynthesis is a process that occurs in an open
quantum system.

We mention here also the paper \cite{lakhno} on the dynamical theory
of primary processes of charge separation in the photosynthetic
reaction center. The emphasis in this paper is on the 
important role of the primary processes, in which light energy is
converted into energy being necessary for the living organisms
to work. The lifetime of the primarily
excited state must be very short. Otherwise
there is no chance for the reaction center to catch the
energy received from the photosynthetic excitation which will 
change, instead, to heat and fluorescence (in the framework of 
Hermitian quantum physics).

In the description of an open quantum system by means of a
non-Hermitian Hamilton operator,
the localized part of the system is embedded into an environment.
Mostly, the environment is the 
extended continuum  of scattering wavefunctions,
see e.g. the review \cite{top}. Coherence is an important ingredient
of this formalism. Meanwhile
the non-Hermitian formalism is applied successfully to the
description of different realistic open quantum systems, see the
recent review \cite{ropp}. 

The paper \cite{klro} is
one of the oldest references in which the 
resonance structure of the cross section in the regime
of overlapping resonances
is considered in the non-Hermitian formalism.
In this paper, the resonance structure of the nuclear
reaction $^{15}N+p$ with two open decay channels is traced
as a function of the degree of overlapping of the individual 
resonances by keeping constant the coupling strength between 
the localized part of the system and the
environment of scattering wavefunctions. The distance between
the energies of the individual resonance states is varied by hand. As
a result, two short-lived states are formed at a critical value of the
degree of overlapping. The widths of all the other states are reduced
because $\sum_{n=1}^N \Gamma_n $ has to be constant according to the
constant coupling strength between system and environment. These
states are called  trapped states. 

In some following papers, this phenomenon is studied as a function of
the coupling strength between system and environment and is called
segregation of decay widths, see the recent review  \cite{au-zel}. In
these papers, the short-living states are called  superradiant
states which exist together with long-living subradiant states. This
formalism is applied also to the problem of energy transfer in 
photosynthetic complexes \cite{celardo1,celardo2}, see also 
\cite{berman1}. In this formalism, the enhancement of the energy 
transfer is related to the existence of the superradiant state.

In other papers, the resonance trapping  phenomenon 
is related to  singular points which exist in the  formalism of
non-Hermitian quantum physics, see the review \cite{top} and
the recent paper \cite{proj10}. These singular points are
known in mathematics since many years \cite{kato}, and are called
usually  exceptional points (EPs).  Most interesting new features
caused by the EPs in the
non-Hermitian physics  of open quantum systems are, 
firstly, the   non-rigid phases of the eigenfunctions and, 
secondly, the possibility of an
external mixing (EM) of the states of the localized part of the
system via the environment. Non-rigidity of the phases of the
eigenfunctions of the Hamiltonian and an EM of the states  are
possible 
only in an open quantum system. They are not involved explicitly in
any type of Hermitian quantum physics. 
Furthermore, superradiant and subradiant states do not appear in this
formalism. Quite the contrary,
phenomena that are related in, e.g., \cite{au-zel} to their existence, 
are an expression for nothing but the nontrivial properties  of the
eigenfunctions of a non-Hermitian Hamilton operator, such as non-rigid
phases and EM of the wavefunctions.

In \cite{nest1-2}, the dynamics of the system and the efficiency  
of energy transfer are studied
in a non-Hermitian formalism  by taking into account 
noise acting between donor and acceptor, while in
\cite{nest3-4}, the
role of protein fluctuation correlations in the energy transfer is
investigated and the spin-echo approach is extended to include
bio-complexes for which the interaction with dynamical noise is strong.

It is the aim of the present paper to provide
the general formalism of
non-Hermitian physics of open quantum systems \cite{top,proj10}
by inclusion of gain which simulates the acceptor, as well as of 
loss which  stands for the donor \cite{comment3}. 
When additionally 
the coupling of the system to a  sink is taken into account,
this formalism can be applied to the description of
light-harvesting of photosynthetic complexes.
We underline that this
formalism describes  the process of photosynthesis as a whole,
i.e. as a  uniform process. While the first part occurs 
instantly, the second part of the process
may last  longer. The formalism 
is generic. In the future, it has to be applied  to concrete
systems with realistic parameters.

In Sect. \ref{gainloss1}, we sketch the formalism for the study of an
open quantum system with gain and loss which is basic 
for the description of photosynthesis. In Sect. \ref{gainloss2}, we
include additionally a sink into the formalism simulated by coupling
to a second environment. 
In both sections we provide analytical as well as numerical results.
We discuss and summarize the results in Sect. \ref{disc}
and draw some conclusions in  Sect. \ref{concl}. 

Before providing the formalism for the description of
open quantum systems, it is necessary to clarify the meaning 
of some terms. We will use definitions similar to those used in
nuclear physics. 

* In nuclear physics, channel  denotes the coupling of a certain  
state of the nucleus $A$ to its decay products after emission of
particle $a$ and  leaving the residual nucleus $(A-a)$ 
in a special state. 
The term channel is equivalent to  embedding of a localized state
 of the system into an environment.
The localized state in nuclear physics is the state of the nucleus
$A$, while the environment is the
 the continuum of scattering wavefunctions of the particle $a$.

* In difference to the definition of energy and width of a nuclear
state  in nuclear physics,
we use the definition $\varepsilon_k = e_k+\frac{i}{2}\gamma_k$
for the complex eigenvalues of the non-Hermitian 
Hamilton operator $\cal H$. The widths of decaying states have thus a
negative sign \cite{comment1}.

* The term internal  mixing of the wavefunctions denotes the
direct interaction between two orthogonal eigenfunctions of a
Hermitian Hamilton operator, 
$\langle \Phi_i|V|\Phi_{j\ne i} \rangle$.
In our calculations, it is supposed to be included in the energies
$e_k$ and widths $\gamma_k$ of the states
that define the non-Hermitian Hamilton matrix, see e.g. Eq. (\ref{gain1}). 
An  external mixing of two eigenstates of a non-Hermitian
Hamilton operator occurs via the environment and is thus a
second-order process. It is defined only in an open system.

* The singularity related to the coalescence \cite{comment2}
of two eigenvalues of a non-Hermitian Hamilton operator $\cal H$  is
called, in recent literature, mostly 
exceptional point. In older papers, the equivalent expressions 
branch point in the complex plane or  double pole of the
S-matrix are mostly used.

\section{Open quantum  systems with gain and loss}
\label{gainloss1}

\subsection{Hamiltonian}
\label{ham1}

We sketch the features characteristic of an open quantum system with
gain and loss \cite{comment3} by considering a localized 2-level system 
that is embedded in a common continuum of scattering wavefunctions. 
One of these two states gains particles from the environment
by interacting with it, while the other one  loses particles to the
continuum by decay.

For the description of the open quantum system, 
we use the non-Hermitian Hamilton operator \cite{proj10}  
\begin{eqnarray}
\label{gain1}
\tilde {\cal H}^{(2,1)} = 
\left( \begin{array}{cc}
\varepsilon_{1}^{(1)} \equiv e_1^{(1)} + \frac{i}{2} \gamma_1^{(1)}  
& ~~~~\omega^{(1)}   \\
\omega^{(1)} 
& ~~~~\varepsilon_{2}^{(1)} \equiv e_2 + \frac{i}{2} \gamma_2^{(1)}   \\
\end{array} \right) \; .
\end{eqnarray}
Here, $\varepsilon_i^{(1)}$ are the two complex eigenvalues of the basic 
non-Hermitian operator coupled to the environment $1$
(called also channel $1$) \cite{comment1}.
The $e_i^{(1)}$ are the  energies of the states $i$ and the
$\gamma_i^{(1)}$ are their widths.
One of these eigenvalues  describes  loss characteristic of
decaying states  ($\gamma_2^{(1)} < 0$) while the other one 
describes  gain from the environment ($\gamma_1^{(1)} > 0 $) 
\cite{comment1}.

The $\omega^{(1)}$ stand for the coupling matrix elements of the two
states via the common environment $1$. They are complex \cite{top}.
The complex eigenvalues ${\cal E}_i^{(1)} \equiv  E_i^{(1)} +
\frac{1}{2} \Gamma_i^{(1)}$ of ${\cal H}^{(2,1)}$ give the energies 
$E_i^{(1)}$ and widths $\Gamma_i^{(1)}$ of the states of the localized
part of the system \cite{comment1}.

We will consider also the non-Hermitian Hamilton operator
\begin{eqnarray}
\label{gain2}
\tilde{\cal H}_0^{(2,1)} = 
\left( \begin{array}{cc}
\varepsilon_{1}^{(1)} \equiv e_1^{(1)} + \frac{i}{2} \gamma_1^{(1)}  & 0   \\
0 & ~~~~\varepsilon_{2}^{(1)} \equiv e_2^{(1)} + \frac{i}{2} \gamma_2^{(1)}   \\
\end{array} \right) 
\end{eqnarray}
which describes the localized part of the open system without coupling
of its states via the continuum ($\omega^{(1)}=0$). 
The phases of the eigenfunctions $\Phi^0_i$ of 
$\tilde{\cal H}_0^{(2,1)} $ are rigid (like in Hermitian
quantum physics) when $\gamma_1^{(1)} = - \gamma_2^{(1)}$.

\subsection{Eigenvalues}
\label{eigenv1}

The eigenvalues of $\tilde \ch^{(2,1)}$ are, generally, complex and may
be expressed as
\begin{eqnarray}
\ce_{1,2}^{(1)} \equiv E_{1,2}^{(1)} + \frac{i}{2} \Gamma_{1,2}^{(1)} = 
\frac{\varepsilon_1^{(1)} + \varepsilon_2^{(1)}}{2} \pm Z ~; \quad \quad
Z \equiv \frac{1}{2} \sqrt{(\varepsilon_1^{(1)} -
  \varepsilon_2^{(1)})^2 + 4 (\omega^{(1)})^2}
\label{int6}
\end{eqnarray}
where  $E_i^{(1)}$ and $\Gamma_i^{(1)}$ stand for the energy and width,
respectively, of the eigenstate $i$. Also here $\Gamma_i^{(1)} \le 0$ for
decaying states and $\Gamma_i^{(1)} \ge 0$ for 
gaining states \cite{comment1}.
The two states may repel each other in accordance with Re$(Z)$,
or they may undergo width bifurcation in accordance with Im$(Z)$.
When $Z=0$ the two states cross each other at a point that 
is called usually  exceptional point (EP) \cite{kato}. The EP is a
singular point (branch point) in the complex plane where the $S$-matrix has 
a double pole \cite{top}. According to its definition 
\cite{kato},  the EP is meaningful in an open quantum system  
which is embedded in  one common environment
$c=1$. Correspondingly, we denote e.g. the eigenvalues by
$\ce_{i}^{(1)}$.

We consider now the behavior 
of the eigenvalues  when the 
parametrical detuning of the two eigenstates of $\tilde\ch^{(2,1)}$ is varied, 
bringing them towards coalescence \cite{comment2}. 
According to (\ref{int6}), the  condition for coalescence reads 
\begin{eqnarray}
Z = \frac{1}{2} \sqrt{(e_1^{(1)}-e_2^{(1)})^2 - 
\frac{1}{4} (\gamma_1^{(1)}-\gamma_2^{(1)})^2 
+i(e_1^{(1)}-e_2^{(1)})(\gamma_1^{(1)}-\gamma_2^{(1)}) +
4(\omega^{(1)})^2} 
~=~ 0 \; .
\label{int6i}
\end{eqnarray}
We consider  two  cases that can be solved analytically.

\begin{enumerate}
\item[(i)]
When $e_1^{(1)} = e_2^{(1)} $, and  $\omega^{(1)} $  is real, it follows 
from (\ref{int6i}) the condition
\begin{eqnarray}
\frac{1}{4}(\gamma_1^{(1)} - \gamma_2^{(1)})^2 =4\, (\omega^{(1)})^2  
~~\rightarrow ~~\gamma_1^{(1)} - \gamma_2^{(1)} =\pm \, 4\, \omega^{(1)} 
\label{int6b}
\end{eqnarray}
for the coalescence of the two eigenvalues, i.e. for an EP. 
It follows furthermore  
\begin{eqnarray}
\label{int6c}
(\gamma_1^{(1)} - \gamma_2^{(1)})^2 <16\, (\omega^{(1)})^2 
&\rightarrow& ~Z ~\in ~\Re \\
\label{int6d}
(\gamma_1^{(1)} - \gamma_2^{(1)})^2 >16\, (\omega^{(1)})^2 
&\rightarrow&  ~Z ~\in ~\Im \; . 
\end{eqnarray}
Eq. (\ref{int6c}) describes the behavior of the eigenvalues 
away from the EP, where the 
eigenvalues ${\cal E}_{k}^{(1)}$  differ from the original ones 
through only a contribution to the energy. 
The widths in contrast, remain
unchanged, and this situation therefore corresponds to that of level
repulsion.
Eq. (\ref{int6d}), in contrast, is relevant 
at the other side of the EP. Here, 
the resonance states  undergo width bifurcation 
according to Im$(Z)\ne 0$.
The bifurcation starts in the neighborhood of the EP.
Physically, the bifurcation implies that  different time scales
may appear in the system, while the states are nearby in energy. 

\item[(ii)] 
When $e_1^{(1)} = - e_2^{(1)}\ne 0$, and  $\omega^{(1)} $  
is imaginary, then the condition
\begin{eqnarray}
(2e^{(1)})^2
=  4\, (\omega^{(1)})^2  
~~\rightarrow ~~
2e^{(1)} =\pm \, 2\, \omega^{(1)} \; ,
\label{int6b1}
\end{eqnarray}
together with $\gamma_1^{(1)} = \gamma_2^{(1)}$,
follows for the coalescence of the two eigenvalues
from (\ref{int6i}).
Here $2e^{(1)} \equiv e_1^{(1)} - e_2^{(1)} $. 
Instead of (\ref{int6c}) and  (\ref{int6d}) we have
\begin{eqnarray}
\label{int6cc}
(2e)^2 > 4\,(\omega^{(1)})^2
&\rightarrow& ~Z ~\in ~\Re \\
\label{int6dd}
(2e)^2 < 4(\omega^{(1)})^2
 &\rightarrow&  ~Z ~\in ~\Im \; . 
\end{eqnarray}
Thus, the EP causes width bifurcation also in this case. However, 
this case is realized 
only when $\gamma_1^{(1)} = \gamma_2^{(1)} = 0$ at the EP,
i.e. when gain and loss vanish at the EP.
\end{enumerate}

\subsection{Eigenfunctions}
\label{eigenf1}

The eigenfunctions of a non-Hermitian Hamilton operator are biorthogonal
(for details see \cite{top,proj10})
\begin{eqnarray}
\label{eif1}
{\cal H} |\Phi_i\rangle =  {\cal E}_i|\Phi_i\rangle \hspace*{1cm}
\langle \Psi_i|{\cal H} = {\cal E}_i \langle  \Psi_i|\; .
\end{eqnarray}
In the case of the symmetric $2\times 2$ Hamiltonian (\ref{gain1}), it is 
\begin{eqnarray}
\label{eif1a}
\Psi_i = \Phi_i^* 
\end{eqnarray}
and the eigenfunctions  should be normalized according to 
\begin{eqnarray}
\label{eif3}
\langle \Phi_i^*|\Phi_j\rangle = \delta_{ij} 
\end{eqnarray}
in order to smoothly describe the transition from a closed system 
with discrete states to a weakly open one with narrow resonance
states. As a consequence of (\ref{eif3}), the values of the 
standard expressions  are changed,
\begin{eqnarray}
\label{eif4}
\langle\Phi_i|\Phi_i\rangle  =  
{\rm Re}~(\langle\Phi_i|\Phi_i\rangle) ~; \quad
A_i \equiv \langle\Phi_i|\Phi_i\rangle \ge 1 
\end{eqnarray}
\vspace*{-1.2cm}
\begin{eqnarray}
\label{eif5}
\nonumber
\langle\Phi_i|\Phi_{j\ne i}\rangle  = 
i ~{\rm Im}~(\langle\Phi_i|\Phi_{j \ne i}\rangle) =
-\langle\Phi_{j \ne i}|\Phi_i\rangle 
\end{eqnarray}
\vspace*{-1.4cm}
\begin{eqnarray} 
\label{eif6}
|B_i^j|  \equiv |\langle \Phi_i | \Phi_{j \ne i}| ~\ge ~0 \; .  
\end{eqnarray}
Furthermore, the
phase rigidity which is a quantitative measure for the biorthogonality 
of the eigenfunctions, 
\begin{eqnarray}
\label{eif7}
r_k ~\equiv ~\frac{\langle \Phi_k^* | \Phi_k \rangle}{\langle \Phi_k 
| \Phi_k \rangle} ~= ~A_k^{-1} \; , 
\end{eqnarray}
is  smaller than 1. Far from an EP, 
$r_k \approx 1$ while it approaches the value 
$r_k =0$ when an EP is approached.

The Hamiltonian (\ref{gain2})
describes the system around the EP without any
mixing of its states via the environment,
since  $\omega^{(1)} =0$ 
corresponds to vanishing EM of the eigenstates. In order to determine 
quantitatively the strength of the EM, 
we present the eigenfunctions
$\Phi_i$ of $\tilde{\cal H}^{(2,1)}$ in the set of
eigenfunctions  $\{\Phi_i^0\}$ of $\tilde{\cal H}_0^{(2,1)}$,
\begin{equation}
\label{eif12}
\Phi_i=\sum \, b_{ij} ~\Phi_j^0 \; ;
\quad \quad b_{ij} = \langle \Phi_j^{0 *} | \Phi_i\rangle \; ,   
\end{equation}
under the condition that the $b_{ij}$ are normalized by 
$\sum_j (b_{ij})^2 = 1$. The coefficients $|b_{ij}|^2$ 
differ from the $(b_{ij})^2$. They contain the
information on the strength of EM via the environment which is 
determined by the value of $\omega^{(1)}$.

For illustration, we consider the EM of the wavefunctions  
$\Phi_1$ and $\Phi_2$ around an EP in the two cases
discussed in Sect. \ref{eigenv1}.
\begin{enumerate}
\item[(i)]
$e_1^{(1)} = e_2^{(1)}$, and  $\omega^{(1)} \in \Re $:
according to (\ref{int6b}),
the strength of the EM via the environment is determined by the
differences  $|\gamma_1^{(1)}|  - |\gamma_2^{(1)}|$ of the widths
(which both have different sign). It
depends thus on the fluctuations of the  $\gamma_i^{(1)}$.

\item[(ii)] 
$e_1^{(1)} = - e_2^{(1)}$,
and $\omega^{(1)} \in \Im $: according to 
(\ref{int6b1}) the strength of the EM is related to
the differences  $|e_1^{(1)}|  - |e_2^{(1)}|$ of the energies, i.e. to the
fluctuations of the $e_i^{(1)}$. This case is however  realized  only
when gain and loss vanish at the EP 
(i.e. $\gamma_1^{(1)} = \gamma_2^{(1)} = 0$ at the EP).
\end{enumerate}

At the EPs, the two corresponding eigenfunctions are not orthogonal. 
Instead
\begin{eqnarray}
\label{sec8}
\Phi_1^{\rm cr} \to ~\pm ~i~\Phi_2^{\rm cr} \; ;
\quad \qquad 
\Phi_2^{\rm cr} \to
~\mp ~i~\Phi_1^{\rm cr}   
\end{eqnarray}  
according to analytical and numerical results 
\cite{ro01,magunov,gurosa,berggren,berggren2}. 
We underline once more that an EP is, according to its definition,
related to the common 
environment in which the system is embedded. In other words, 
it is well defined under the condition that the system is 
embedded in  {\it only one} continuum.

\subsection{Schr\"odinger equation with source term}
\label{schr1}

The Schr\"odinger equation $(\tilde{\cal H}^{(2,1)}
- {\cal{E}}_i^{(1)} |\Phi_i^{(1)}  \rangle =0$
may be rewritten into a Schr\"odinger equation with source term
\cite{top,proj10},
\begin{eqnarray}
\label{eif11}
(\tilde{\cal H}^{(2,1)}_0  - {\cal E}_i^{(1)}) ~| \Phi_i^{(1)} \rangle  
= -\left(
\begin{array}{cc}
0 & \omega^{(1)} \\
\omega^{(1)} & 0 
\end{array} \right) |\Phi_i^{(1)} \rangle \; . 
\end{eqnarray}
In this representation, 
the coupling $\omega^{(1)}$ of the  states  $i$ and ${j\ne i}$ 
of the localized system
via the common environment of scattering wavefunctions (EM)
is contained solely in the source term.
The source term vanishes, when $e_1^{(1)} = e_2^{(1)}$
around the EP under the condition
$\gamma_1^{(1)} =  \gamma_2^{(1)}$ according to (\ref{int6b}),
what is fulfilled when $\gamma_{i=1,2}^{(1)} = 0$ .

Far from EPs, the coupling of the localized system to
the environment influences the spectroscopic properties of the system, 
in general,
only marginally \cite{top,proj10}. The influence is however 
non-vanishing also in this case, see e.g. the experimental results 
\cite{savin2}.

In the neighborhood of EPs, however, the coupling between 
system and environment and therewith the source term 
play an important role for the dynamics of the open quantum
system. The reason is, according to mathematical 
studies, that the source term  causes 
nonlinear effects in the Schr\"odinger equation (\ref{eif11}) around 
an EP.  For details see \cite{top,proj10}.

\subsection{Resonance structure of the S-matrix}
\label{smatr1}

Let us consider the  resonance part of the $S$ matrix
from which the resonance structure 
of the cross section can be calculated, 
\begin{eqnarray}
\label{cro}
\sigma (E) \propto |1-S(E)|^2 \; .
\end{eqnarray}
A unitary representation of the resonance part of the 
$S$ matrix in the case of two resonance states coupled to a 
common continuum of scattering wavefunctions reads \cite{ro03} 
\begin{eqnarray}
\label{sm4}
S = \frac{(E-E_1-\frac{i}{2}\Gamma_1)~(E-E_2-\frac{i}{2}\Gamma_2)}{(E-E_1+
\frac{i}{2}\Gamma_1)~(E-E_2+\frac{i}{2}\Gamma_2)}\; .
\end{eqnarray}
Here, the influence of the EPs onto the cross section is contained 
in the eigenvalues 
${\cal{E}}_i = E_i + i/2~\Gamma_i$. The expression (\ref{sm4})
allows therefore to receive  reliable results  also when the phase 
rigidity is reduced, $r_k < 1$.

Let us assume real $\omega^{(1)}$ and $\gamma_1^{(1)} = - \gamma_2^{(1)}$. 
First we consider
the case corresponding to the condition 
(\ref{int6c}), i.e. for large coupling strength $\omega^{(1)}$  of the 
system to the environment of continuous scattering wavefunctions.
In this case $\Gamma_1 = - \Gamma_2 =0$ and, according 
to (\ref{int6c}),  
\begin{eqnarray}
\label{sm14}
S = \frac{(E-E_1)~(E-E_2)}{(E-E_1)~(E-E_2)} =1 \; .
\end{eqnarray}
In the other case,  (\ref{int6d}), it is  $E_1=E_2$; 
$\Gamma_1=-\Gamma_2 \ne 0$; and 
\begin{eqnarray}
\label{sm24}
S = \frac{(E-E_1-\frac{i}{2}\Gamma_1)~(E-E_1+\frac{i}{2}\Gamma_1)}{(E-E_1+
\frac{i}{2}\Gamma_1)~(E-E_1-\frac{i}{2}\Gamma_1)} =1\; .
\end{eqnarray} 
In both cases, $S=1$, i.e. $\sigma(E) =0$ according to (\ref{cro}).
This result corresponds to the well-known fact that EPs cannot be
identified in the resonance structure of the S-matrix and therefore 
also not in the resonance structure of the cross section.
Most important is however the result that no resonances will be 
excited due to $\gamma_1^{(1)} = - \gamma_2^{(1)}$.

The result $S=1$ is violated  when the conditions 
$\omega^{(1)} \in \Re $ 
and $\gamma_1^{(1)} = - \gamma_2^{(1)}$ are not exactly 
fulfilled. This may happen, e.g., under the influence of external 
random (stochastic) processes that cause fluctuations of the
$\gamma_i^{(1)}$. In such a case, $S<1$; and the energy (or
information) will be transferred with an
efficiency of nearly 100 \% (because no resonances can be excited
under this condition in the localized part of the system).
Results for this case can be obtained only numerically.

\subsection{Numerical results: one-channel case}
\label{num1}

\begin{figure}[ht]
	\begin{center}
\includegraphics[width=12.cm,height=11.5cm]{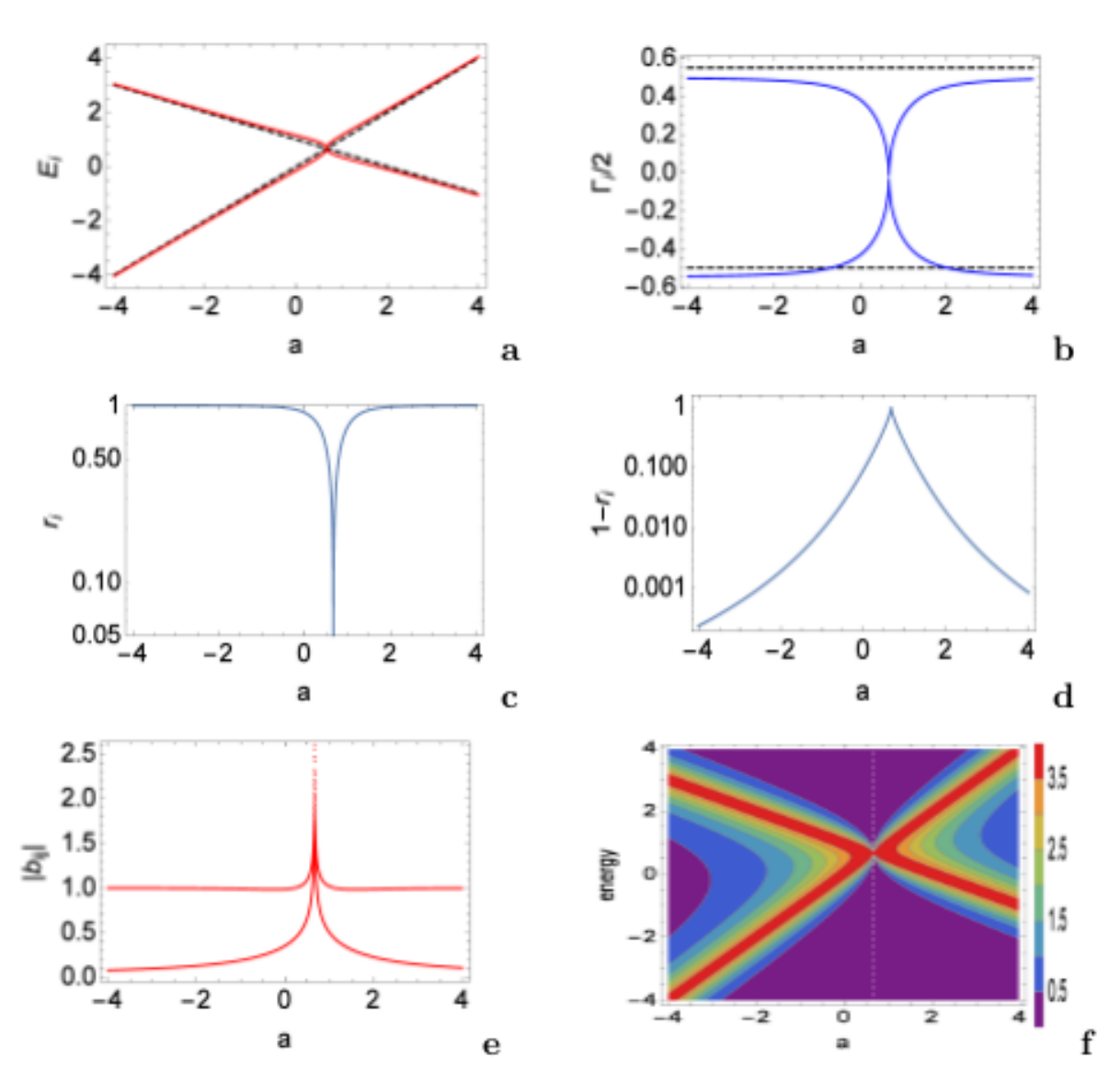}
		\vspace*{-.7cm} 
	\end{center}
	\caption{
		\footnotesize{
			Eigenvalues ${\cal E}_i^{(1)} \equiv E_i^{(1)} +
			\frac{i}{2}\Gamma_i^{(1)}$ (a,b)
			and eigenfunctions $\Phi_i^{(1)}$ (c,d,e)
			of the Hamiltonian $\tilde{\cal H}^{(2,1)}$ 
			as a function of $a$ and  contour plot (f) of
                        the cross section; ~$\omega^{(1)} = 0.5025$.
                        Parameters: $e_1^{(1)}= 1-a/2; ~e_2^{(1)}=a; 
			~\gamma_1^{(1)}/2=0.55; ~\gamma_2^{(1)}/2=-0.5$
                        (dashed lines in a, b).		}} 
		\label{fig1}
	\end{figure}

\begin{figure}[ht]
	\begin{center}
\includegraphics[width=12.cm,height=11.5cm]{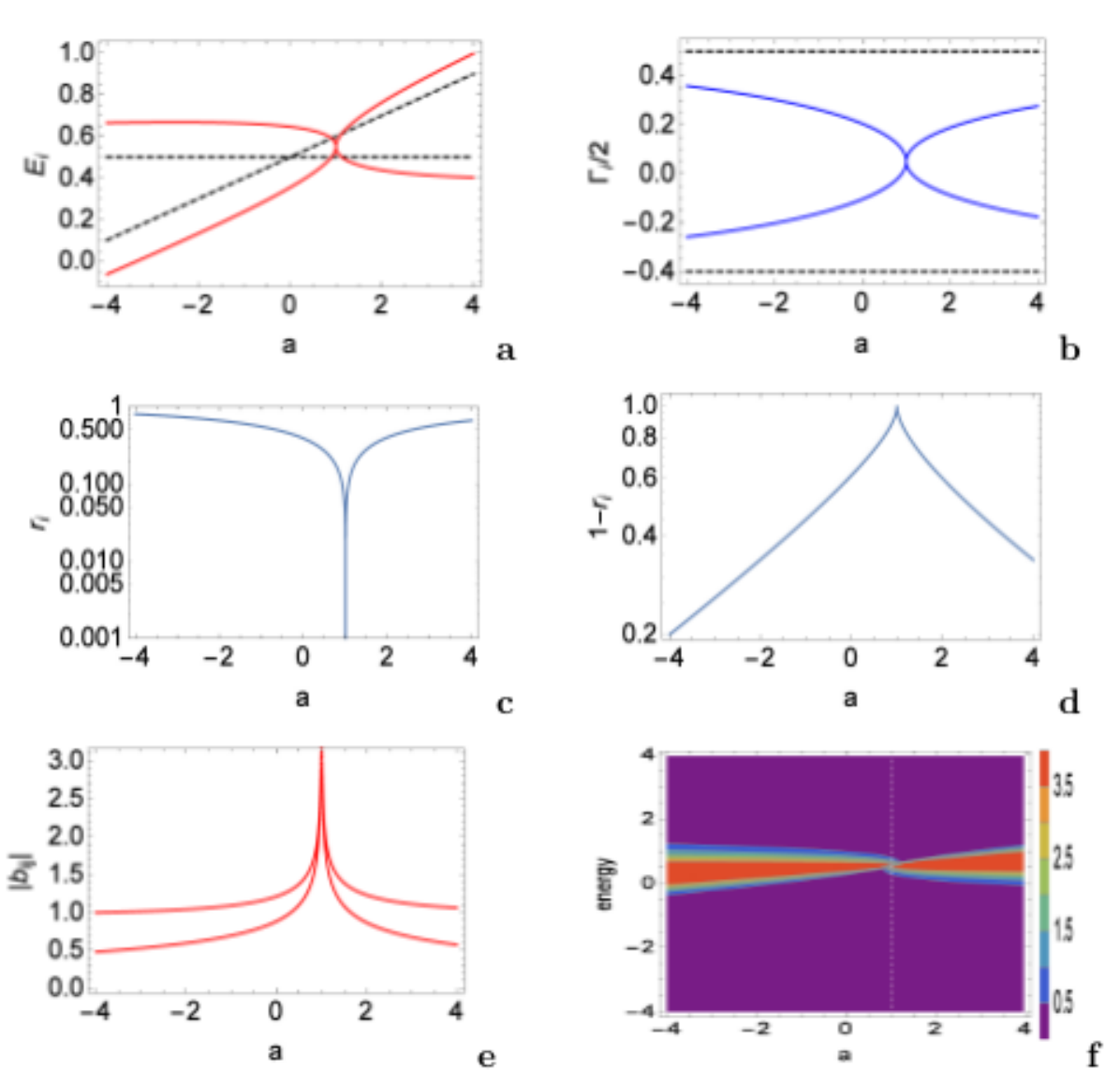}
		\vspace*{-.7cm} 
	\end{center}
	\caption{
		\footnotesize{
The same as Fig. \ref{fig1} but                         
$\omega^{(1)} = 0.45+0.05i$ and  $e_1^{(1)}= 0.5; ~e_2^{(1)}=0.5+0.1a; 
~\gamma_1^{(1)}/2=0.5; ~\gamma_2^{(1)}/2=-0.4$.
}} 
		\label{fig2}
	\end{figure}

\begin{figure}[ht]
	\begin{center}
\includegraphics[width=12.cm,height=11.5cm]{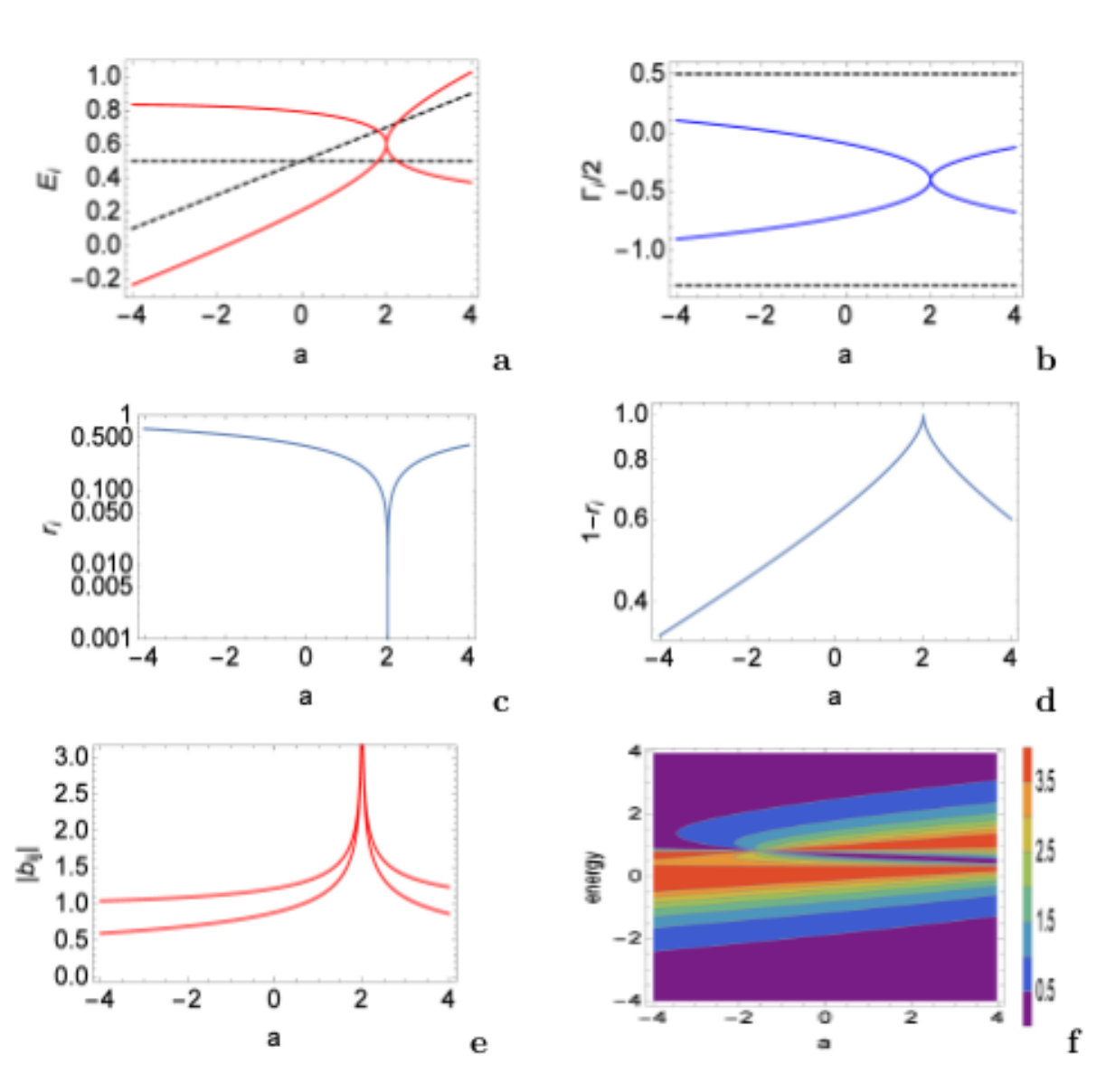}
		\vspace*{-.7cm} 
	\end{center}
	\caption{
		\footnotesize{
The same as Fig. \ref{fig1} but
$\omega^{(1)} = 0.9+0.1i$ and  
$e_1^{(1)}= 0.5; ~e_2^{(1)}=0.5+0.1a; 
~\gamma_1^{(1)}/2=0.5; ~\gamma_2^{(1)}/2=-1.3$.
}} 
		\label{fig3}
	\end{figure}

\subsubsection{Merge of states with gain and loss}
\label{num1a}

Let us first consider the results that are obtained by using a
parametric dependence of the energies $e_i^{(1)}$
and widths $\gamma_i^{(1)}$ of the states of the localized part of the
system which is analog to that
used in \cite{epj1,proj10} for a decaying system. 
The results are shown in Fig. \ref{fig1}.

In this figure, the existence of an
EP at the parameter value  $a=a^{\rm cr} = 2/3$
can clearly be seen. Here, the two states are exchanged
(Figs. \ref{fig1}.a,b),  the phase rigidity
$r_i$ approaches the value $0$ (Figs. \ref{fig1}.c,
d) and the EM 
of the states via the continuum increases limitless
(Fig. \ref{fig1}.e). The contour plot of the cross section
(Fig. \ref{fig1}.f)
shows the wavefunctions of the two states: while the eigenvalues of 
$\tilde {\cal H}^{(2,1)}$ cannot be seen according to (\ref{sm14}) and 
(\ref{sm24}), the eigenfunctions show some fluctuating behavior around
the positions of the eigenstates according to the finite
(nonvanishing) range of their influence (see
Figs. \ref{fig1}.c,d,e). These fluctuations of the eigenfunctions can
be seen in the contour plot. Although they follow the positions of the
eigenvalues, their nature is completely different from that of the
eigenvalue trajectories. The eigenfunction trajectories in
Fig. \ref{fig1}.f show
the exchange of the two states at the EP. That means: the state
with positive width  turns into a state with negative width and vice
versa. This underlines once more that the two trajectories shown in
Fig. \ref{fig1}.f  have really nothing in common with  the 
eigenvalue trajectories of resonance states the
widths of which are always negative (or zero at most).
   
We underline once more that the results shown in Fig. \ref{fig1} are  
formally similar to those obtained and
discussed in \cite{epj1,proj10} for a decaying system.
In the latter case,  both states which are exchanged at the EP,  are
of the same type: they are resonance states with negative widths.
It is interesting to see from the numerical results (Fig. \ref{fig1}), 
that the non-Hermitian Hamilton operator $\cal H$ can be used, indeed, 
for the description of these two different  types of 
open quantum systems as suggested in Sect. \ref{gainloss1} 
\cite{comment3}. 

Additionally we show some numerical results for the case that, 
respectively, the distance in energy of the two states 
is smaller (Fig. \ref{fig2}) and the EM as well as
the widths  $|\gamma_i^{(1)}|$ differ more 
from one another (Fig. \ref{fig3}) than in
Fig. \ref{fig1}. The eigenvalue and eigenfunction figures 
(Figs. \ref{fig1}.a-e,  ~\ref{fig2}.a-e,  
~\ref{fig3}.a-e) are similar to one another and show clearly the
signatures of an EP at a certain critical value of the parameter
$a=a_{\rm cr}$. The contour plots  
(Fig. \ref{fig1}.f, ~\ref{fig2}.f,  ~\ref{fig3}.f) differ however 
from one another. 

When the states are nearer to one another in energy, the two states
with negative and positive width merge
(Fig. \ref{fig2}.f).  Under the influence of stronger EM (stronger 
coupling strength $\omega^{(1)}$ between system and environment) 
as well as of a larger difference between the two values 
$|\gamma_i^{(1)}|$, the extension of the region with non-vanishing 
cross section is enlarged in relation to the energy (Fig.
\ref{fig3}.f). In any case, the cross section vanishes around 
$a=a_{\rm cr}$. Resonance states are not excited.

\subsubsection{Level repulsion of states with gain and  loss}
\label{num1b}

More characteristic for an open quantum system with gain and loss
than those in Sect. \ref{num1a} are 
the analytical results given in 
Sect. \ref{smatr1}. According to these results, the 
cross section is zero when  $\gamma_1^{(1)} = - \gamma_2^{(1)}$, 
~$e_1^{(1)} = e_2^{(1)}$, and  $\omega^{(1)} \in \Re $.
Under the influence of an EP which causes differences between
the original spectroscopic values  $\varepsilon_i^{(1)}
\equiv e_i^{(1)} + \frac{i}{2} \gamma_i^{(1)}$
and the eigenvalues ${\cal{E}}_i^{(1)}
\equiv E_{i}^{(1)} + \frac{i}{2} \Gamma_{i}^{(1)} $ of
$\tilde \ch^{(2,1)}$, 
a non-vanishing cross section is  expected  when at 
least one of the conditions 
$\Gamma_1^{(1)} = - \Gamma_2^{(1)}$, 
~$E_1^{(1)} = E_2^{(1)}$, together with  $\omega^{(1)} \in \Re $,
is not fulfilled.

\begin{figure}[ht]
	\begin{center}
\includegraphics[width=12.cm,height=11.5cm]{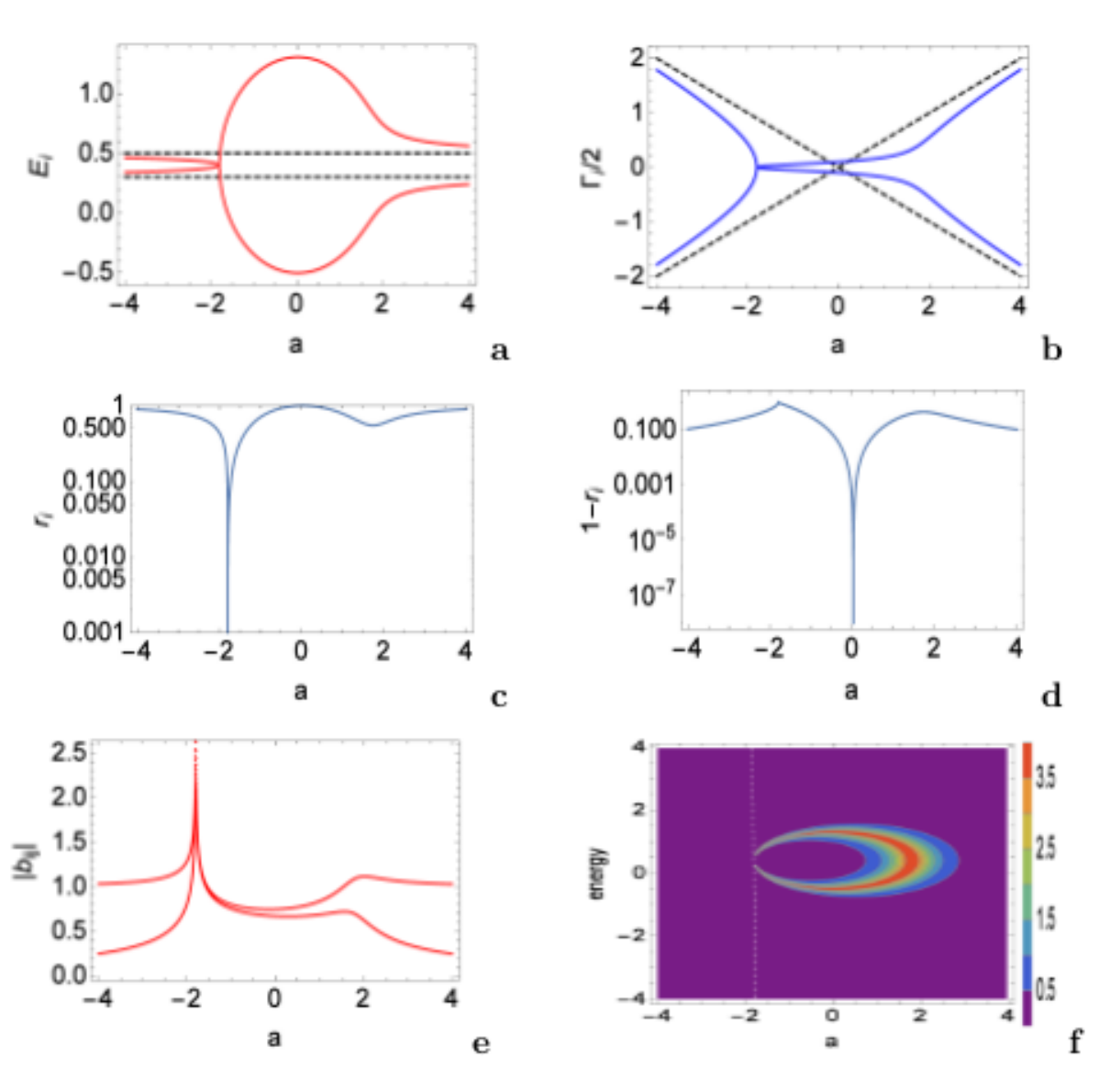}
		\vspace*{-.7cm} 
	\end{center}
	\caption{
		\footnotesize{
The same as Fig \ref{fig1} but 
$\omega^{(1)} = 0.9+0.1i$ and  $e_1^{(1)}= 0.5; ~e_2^{(1)}=0.3; 
			~\gamma_1^{(1)}/2=0.5a; ~\gamma_2^{(1)}/2=-0.5a$.
}} 
		\label{fig4}
	\end{figure}

In Fig. \ref{fig4} we show the corresponding  numerical results 
obtained for two neighboring states, 
$e_1^{(1)} \approx e_2^{(1)}$, and  $\omega^{(1)}$ almost real.
We fix the energies  $e_i^{(1)} $ and vary parametrically the
widths $\gamma_i^{(1)}$, see the dashed lines in Figs. \ref{fig4}.a,b.  
The results show an EP at $a=-1.8$ and the hint to another EP at 
$a=1.8$. At the EP, the phase rigidity approaches 
the value zero, 
$r_i \to 0$ (Fig. \ref{fig4}.c), and the EM of the states is extremely
large, $|b_{ij}| \to \infty$ (Fig. \ref{fig4}.e).

Of special interest is the parameter range between the two EPs.
Here,  $r_i \to 1$ at $a\approx 0$  (Figs. \ref{fig4}.c,d). 
At this parameter value, the level repulsion is maximum; 
and the two eigenfunctions of the non-Hermitian Hamiltonian 
$\tilde \ch^{(2,1)}$ are (almost) orthogonal.

An analogous result is known  from calculations for decaying systems, 
i.e. for systems with excitation of resonance states \cite{top,proj10}.
In these calculations, the energies are varied parametrically
and $\omega$ is almost imaginary (Fig. 1 in \cite{proj10}). Therefore,
the two eigenfunctions of the non-Hermitian Hamilton operator are
(almost) orthogonal ($r_i \to 1$)  at maximum width bifurcation 
(instead of at maximum level repulsion  in Fig. \ref{fig4}).

In any case, the
two eigenstates of the non-Hermitian operator  turn irreversibly into
two states with rigid phases in spite of the
non-Hermiticity of the Hamiltonian. This unexpected result  
occurs due to the evolution  
of the system to the point of, respectively, maximum level repulsion
and maximum width bifurcation,  which is
driven exclusively by the nonlinear  source term  
of the Schr\"odinger equation, see Sect. \ref{schr1}.
The eigenfunctions of these two states are mixed.   

In order to receive a better understanding of this result, we
mention here another unexpected result of non-Hermitian quantum 
physics, namely the fact that a non-Hermitian Hamilton
operator may have real eigenvalues
\cite{comment3}. This fact is very well known in
literature for a long time, for references see the review \cite{top}.
The corresponding states are called usually  bound states in 
the continuum. 

Most interesting for a physical system is the contour plot of the 
cross section (Fig. \ref{fig4}.f). According to the analytical results 
discussed in Sect. \ref{smatr1},
the cross section vanishes far from the parameter range that is
influenced by an EP. It does however not vanish 
completely in the parameter range
between the two EPs. Around the EP at  $ a=-1.8$, the 
conditions for vanishing cross section are quite well fulfilled, while
this is not the case around the other EP at $ a = 1.8$.
In approaching the two EPs  
by increasing and decreasing, respectively, the value of $a$
the cross section vanishes around  $ a=-1.8$,
and is non-vanishing around $ a=1.8$. The cross section vanishes 
also around the point of maximum level repulsion
 at which the two eigenfunctions of $\tilde \ch^{(2,1)}$
 are orthogonal (and {\it not} biorthogonal). 

Additionally, we performed calculations (not shown in the paper) with 
the reduced value $\omega^{(1)}_{red} = 0.5(0.9+0.1i)$ in 
order to determine the role of EM in the cross section picture.
The obtained results are similar to those shown in
Figs. \ref{fig4}.a-f. 
The parameter range influenced by the two EPs is however smaller when
$\omega^{(1)} $ is reduced: it ranges from  $a\approx -1$ to 
$a\approx 1$ when $\omega^{(1)} = \omega^{(1)}_{red}$. Accordingly,
the region of the non-vanishing cross section in the contour plot
shrinks in relation to $a$, and also in relation to the energy.
In calculations with vanishing external mixing ($\omega^{(1)} =0$),
the cross section vanishes everywhere.

\section{Open quantum system with gain and loss 
coupled to two environments}
\label{gainloss2}

\subsection{Hamiltonian for coupling to two environments}
\label{ham2}

Let us consider the $4\times 4$ non-Hermitian matrix
\begin{eqnarray}
\label{ham22}
{\tilde{\cal H}^{(2,2)}} = 
\left( \begin{array}{cccc}
\varepsilon_{1}^{(1)} 
 &
~~~~\omega^{(1)} & 0 & 0  \\
\omega^{(1)} & ~~~~\varepsilon_{2}^{(1)}  
& 0 & 0 \\
0 & 0 & \varepsilon_{1}^{(2)}  
 &
~~~~\omega^{(2)}  \\
0 & 0 &
\omega^{(2)} & ~~~~\varepsilon_{2}^{(2)} 
  \\
\end{array} \right) \; .
\end{eqnarray}
Here, $\varepsilon_i^{(1)}\equiv e_i^{(1)} + \frac{i}{2}
\gamma_i^{(1)} $ and 
$\varepsilon_i^{(2)} \equiv e_i^{(2)} + \frac{i}{2} \gamma_i^{(2)}$
are the complex eigenvalues of the basic 
non-Hermitian operator $\tilde{\cal H}^{(2,2)}$ relative to channel $c=1$
and $c=2$, respectively \cite{comment1}.
The two channels (environments) 
are independent of and orthogonal to one another
what is expressed by the zeros in the matrix 
(\ref{ham22}).  One of the channels may be related to 
gain and loss \cite{comment3} (acceptor and donor) considered 
in the previous section \ref{gainloss1}, while the other channel
may simulate a sink. In this case, the two 
widths $\gamma_1^{(1)}$ and $\gamma_2^{(1)}$
have  different sign
relative to the first channel.
Relative to the second channel however,
both $\gamma_i^{(2)}$ are negative according to a usual decay process 
of a resonance state. 

The $\omega^{(1)}$ and $\omega^{(2)}$  stand for the coupling matrix
elements between the two
states $i=1, ~2$ of the localized part of the open quantum system and 
the  environment $c=1$ and $2$, respectively.
In the case considered above, these
two environments are completely different
from one another and should never be related to one another. In more 
detail: an EM of the considered states may be caused {\it only 
by $\omega^{(1)}$  or by   $\omega^{(2)}$}, and never by both
values at the same time \cite{comment5}. This is guaranteed when
$\omega^{(2)}=0$ what is fulfilled
when there is only one state in the second channel. When there are
more states, then  $|\omega^{(2)}|$ should be much smaller than
$|\omega^{(1)}|$ (here we point to the general result that the values
$\omega$ are related to the
widths $\gamma_i$ of the states \cite{top}).

The values $|\gamma_i^{(1)}|$ and $|\gamma_i^{(2)}|$ are independent
of one another and express the different time scales characteristic of
the two channels. 
While the $|\gamma_i^{(1)}|$ will be usually very large, 
the $|\gamma_i^{(2)}|$ are generally much smaller. 
Accordingly, the two-step process as a whole will show,
altogether, a bi-exponential decay: first the decay occurs due to the
exponential quick process; somewhere at its tail it will
however switch over into the  exponential decay of the slow process. 

The Hamiltonian  which describes vanishing coupling
of the states of the localized part of the open quantum system
to both environments is
\begin{eqnarray}
\label{ham220}
\tilde{\cal H}_0^{(2,2)} = 
\left( \begin{array}{cccc}
\varepsilon_{1}^{(1)} 
 &0 & 0 & 0  \\
0 & ~~~~\varepsilon_{2}^{(1)}  
& 0 & 0 \\
0 & 0 & \varepsilon_{1}^{(2)}  
 & 0  \\
0 & 0 &0 & ~~~~\varepsilon_{2}^{(2)} 
  \\
\end{array} \right) 
\end{eqnarray}  
by analogy to (\ref{gain2}).
It does not contain any EM  via an environment.

\subsection{Eigenvalues and eigenfunctions of $\tilde{\cal H}^{(2,2)} $}
\label{eigenvf}

The eigenvalues  ${\cal E}_i^{(c)} \equiv E_i^{(c)}
+\frac{i}{2}\Gamma_i^{(c)}$   and eigenfunctions $\Phi_i^{(c)}$ 
of (\ref{ham22}) are characterized by two numbers: the number $i$  of 
the state ($i=1, 2$) of the localized part of the system and the
number $c$  of the channel ($c=1, 2$), called environment, in which the
system is embedded. Generally,  $E_i^{(1)} \ne E_i^{(2)}$ and 
$\Gamma_i^{(1)} \ne \Gamma_i^{(2)}$.
Also the wave functions $\Phi_i^{(1)}$ and $\Phi_i^{(2)}$ 
differ from one another due to the EM of the 
eigenstates via the environment $c=1$ and $c=2$, respectively.
From a mathematical point of view,
the system has therefore  four states.

An EP influences the dynamics of the 
open quantum system also in the two-channel case.
Without an EP in the considered parameter range in relation to both
channels, we have   $E_i^{(1)} \approx E_i^{(2)}$,
$\Gamma_i^{(1)} \approx \Gamma_i^{(2)}$ and 
$\Phi_i^{(1)} \approx \Phi_i^{(2)}$.
Accordingly, one has to consider effectively only two states 

Under the influence of an EP relative to $c=1$ (or/and relative to
$c=2$), the eigenvalues and eigenfunctions  will be, however,    
different from one another,
$E_i^{(1)} \ne E_i^{(2)}$,  $\Gamma_i^{(1)} \ne \Gamma_i^{(2)}$ and 
$\Phi_i^{(1)} \ne \Phi_i^{(2)}$
in the corresponding parameter range. We have to
consider therefore effectively four states in this case.

According to \cite{kato} an EP is defined  when the system is 
embedded in  one common environment. Under this condition,  
it  causes nonlinear processes in a physical system, which
is the crucial factor for  the dynamical properties of 
an open quantum system \cite{top}. This is valid not only for systems    
all states of which decay (corresponding to some  loss),
but also for systems with loss  and gain, as shown in 
\cite{top}.

Due to the nonlinear processes occurring near to an EP, it is
difficult to receive analytical solutions for the eigenvalues and
eigenfunctions of (\ref{ham22}). We will provide the results of some
numerical simulations, above all with 
$e_1^{(1)} \approx e_2^{(1)}$, almost real  $\omega^{(1)} $
and almost imaginary  $\omega^{(2)} $,  which
is the most interesting and general case for a system with gain and
loss that is coupled to a sink
(see the analytical results obtained with 
$e_1^{(1)} = e_2^{(1)}$ and $\omega^{(1)} \in \Re $,
Eq. (\ref{int6b}), and the corresponding 
results for decaying systems in \cite{proj10}).

\subsection{Schr\"odinger equation with source term
	and coupling to two environments}
\label{schr2}

Using (\ref{ham22}), we can write down the Schr\"odinger equation with 
source term for the two-channel case in analogy to  (\ref{eif11})
for the one-channel case. The corresponding equation reads
\begin{eqnarray}
\label{eif11s4}
({\cal H}^{(2,2)}_0  - {\cal E}_i^{(c)}) ~| \Phi_i^{(c)} \rangle  = -
\left(
\begin{array}{cccc}
0 & \omega^{(1)} & 0 & 0\\
\omega^{(1)} & 0 & 0 & 0\\
0 & 0 & 0 & \omega^{(2)}\\
0 & 0 & \omega^{(2)} & 0
\end{array} \right) |\Phi_i^{(c)} \rangle \; . 
\end{eqnarray}
The source term depends on the coupling of   the system to both
channels, i.e. on  $\omega^{(1)}$ and  on $\omega^{(2)}$. 
We will consider the  general  case with two channels (two
environments) in which $|\omega^{(2)}| \ll |\omega^{(1)}|$. 

We repeat here that, according to their definition \cite{kato},
EPs occur only in the one-channel case, i.e. only 
in the submatrices related either to channel $1$  or to channel $2$.
They are not defined in the $4\times 4$ matrix 
(\ref{ham22}). However, each  EP in one of the two submatrices in
(\ref{ham22}) influences the dynamics of the open two-channel 
system. This will be shown in the following section
by means of numerical results for the case that there is an EP in the 
first channel which simulates acceptor and  donor (gain and loss), 
while the second channel being of standard type with resonance states,
may or may not have an EP.

\subsection{Numerical results: two-channel case}
\label{num2}

\begin{figure}[ht]
	\begin{center}
	\includegraphics[width=12.cm,height=11.5cm]{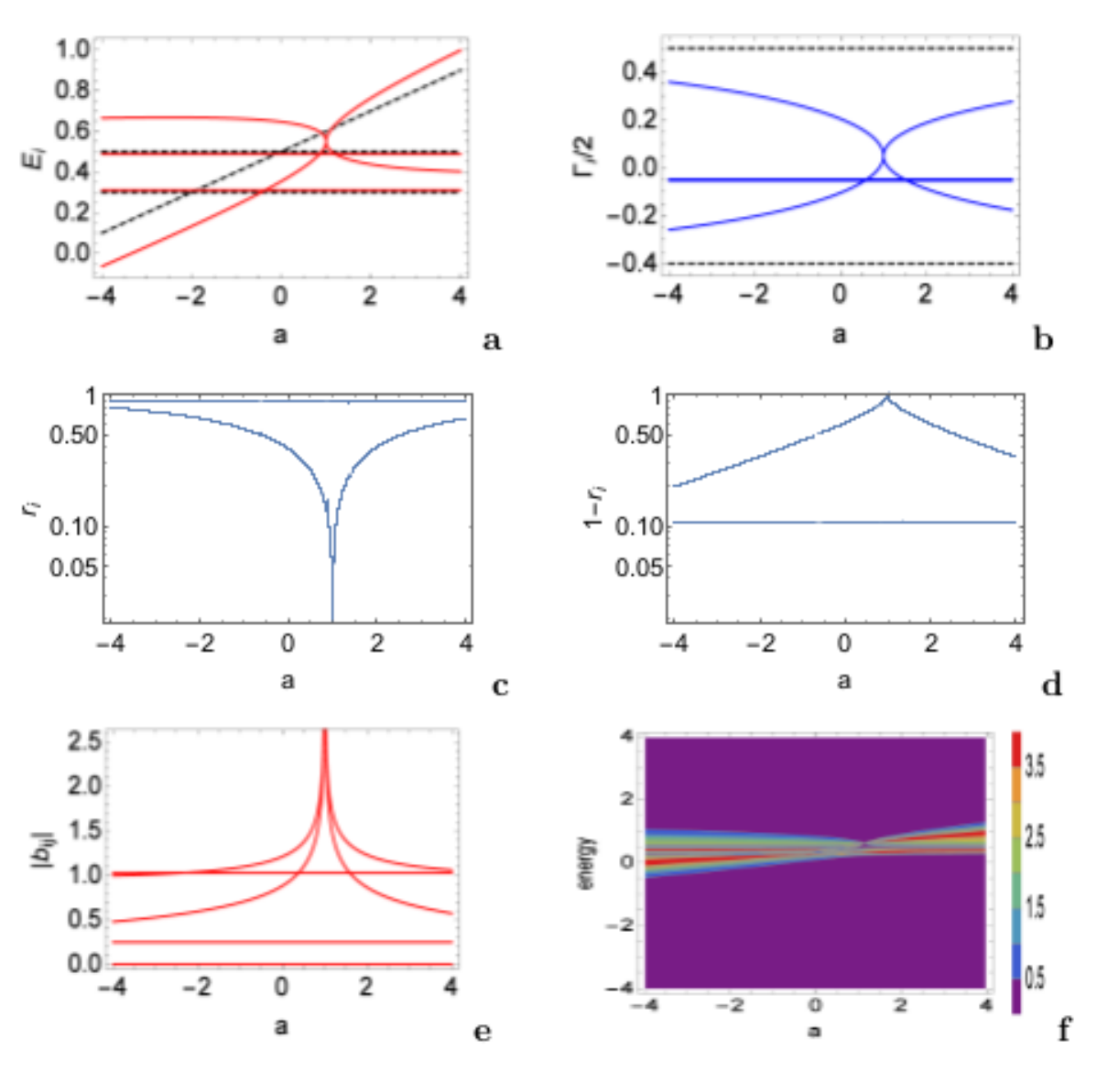}
		\vspace*{-.7cm} 
	\end{center}
	\caption{
		\footnotesize{
	Eigenvalues ${\cal E}_i^{(1,2)} \equiv E_i^{(1,2)} +
			\frac{i}{2}\Gamma_i^{(1,2)}$ (a,b)
			and eigenfunctions $\Phi_i^{(1,2)}$ (c,d,e)
			of the Hamiltonian $\tilde{\cal H}^{(2,2)}$ 
                        and  contour plot (f) of
                        the cross section
			as function of $a$ with two merging
                        states in the
                        first channel, $\omega^{(2)}=0.005+0.045 ~i$.
                        $~e_1^{(2)} = 0.5; ~e_2^{(2)} = 0.3;
                        ~\gamma_1^{(2)}/2=-0.05; ~\gamma_2^{(2)}/2=-0.05$
			(dashed lines in a, b).	
			The parameters of the first channel are the
                        same as those in Fig. \ref{fig2}.
	}}	
		\label{fig5}
	\end{figure}

\begin{figure}[ht]
	\begin{center}
	\includegraphics[width=5cm,height=14cm]{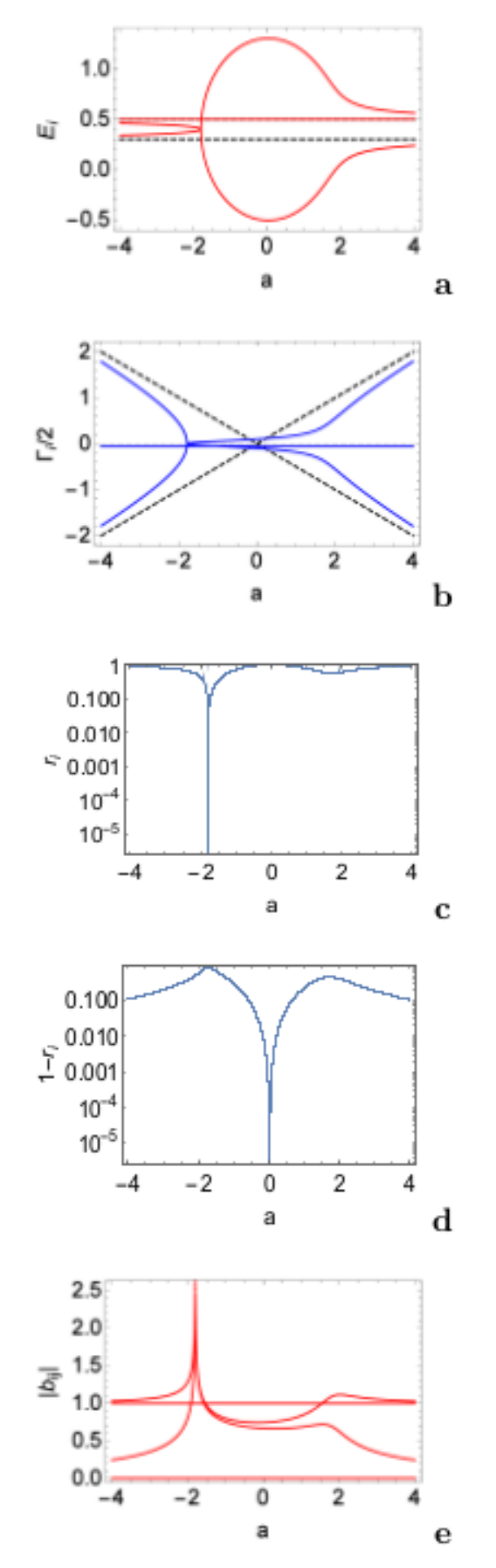}
\hspace*{.2cm} 
\includegraphics[width=5.3cm,height=14.1cm]{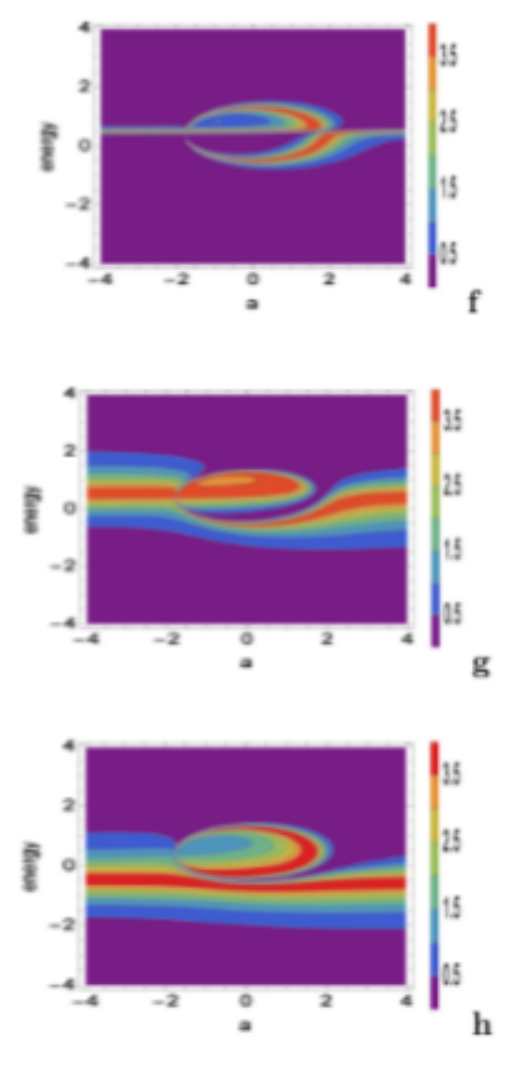}
		\vspace*{-.7cm} 
	\end{center}
	\caption{
		\footnotesize{
 		Left:	Eigenvalues ${\cal E}_i^{(1,2)} \equiv E_i^{(1,2)} +
			\frac{i}{2}\Gamma_i^{(1,2)}$ (a,b)
			and eigenfunctions $\Phi_i^{(1,2)}$ (c,d,e)
			of the Hamiltonian $\tilde{\cal H}^{(2,2)}$ 
			as function of $a$ with one state in the
                        second channel, $\omega^{(2)}=0$.
                        $~e_1^{(2)} = 0.5; ~\gamma_1^{(2)}/2=-0.05$
			(dashed lines in a, b).	
			The parameters of the first channel are the
                        same as those in Fig. \ref{fig4}.
               Right: contour plots of the cross section with the same
                      parameters of the first channel 
                      as  in Fig. \ref{fig4};
                      $\omega^{(2)}=0$; and
                      (f)  $~e_1^{(2)} = 0.5; ~\gamma_1^{(2)}/2=-0.05$;
                    (g)  $~e_1^{(2)} = 0.5; ~\gamma_1^{(2)}/2=-0.5$;
                    (h)  $~e_1^{(2)} =- 0.5; ~\gamma_1^{(2)}/2=-0.5$.
   }}	
	\label{fig6}
\end{figure}

\begin{figure}[ht]
	\begin{center}
\includegraphics[width=12.cm,height=11.5cm]{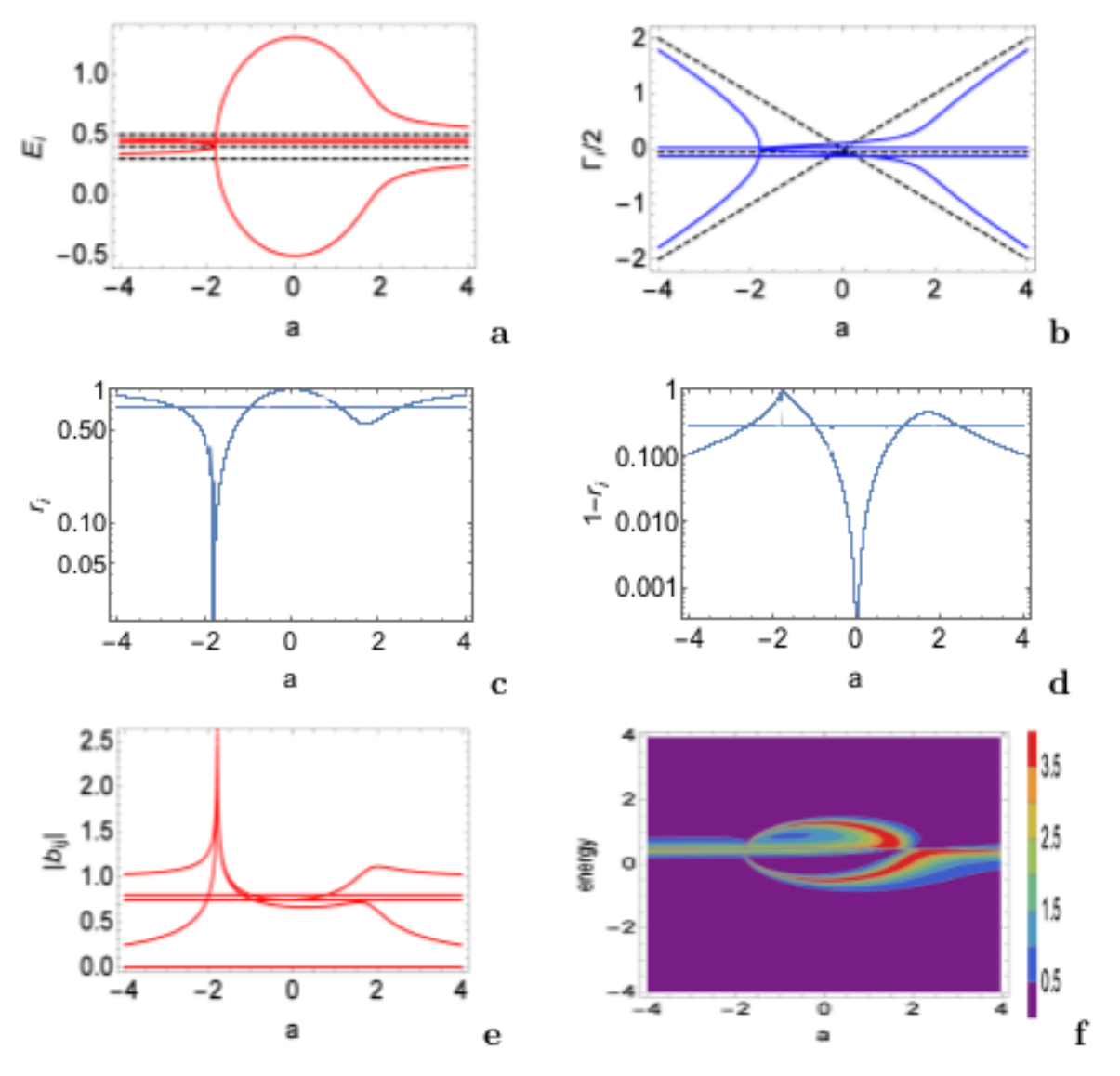}
		\vspace*{-.7cm} 
	\end{center}
	\caption{\footnotesize{
                        Eigenvalues ${\cal E}_i^{(1,2)} \equiv E_i^{(1,2)} +
			\frac{i}{2}\Gamma_i^{(1,2)}$ (a,b)
			and eigenfunctions $\Phi_i^{(1,2)}$ (c,d,e)
			of the Hamiltonian $\tilde{\cal H}^{(2,2)}$ 
                        and  contour plot (f) of
                        the cross section
                        as function of $a$ with two states in the
                        second channel, $\omega^{(2)}=0.01+0.09~i$.
                        $~e_1^{(2)} = 0.4; ~e_2^{(2)} = 0.5;
                        ~\gamma_1^{(2)}/2=-0.06; ~\gamma_2^{(2)}/2=-0.06$
			(dashed lines in a, b).	
			The parameters of the first channel are the
                        same as those in Fig. \ref{fig4}.	
 }} 
		\label{fig7}
	\end{figure}

We performed some calculations for the two-channel case by
starting from the calculations for the one-channel case in,
respectively, Fig. \ref{fig2} and  \ref{fig4}
and by adding a second channel that describes 
decaying states (corresponding to loss). There are, of course, very
many possibilities for choosing the number of states 
as well as the parameters for the second
channel. One possibility is to keep the parameters constant by varying the
parameter $a$ of the first channel. Another possibility is to relate
them directly to the parameter $a$, or to introduce another
independent parameter $b$. The choice should correspond to the
physical situation considered.
 
The aim of our calculations is to illustrate the influence of a second
channel onto the eigenvalues and eigenfunctions of
$\tilde{\cal H}^{(2,2)} $ and onto the contour plot of the cross
section. We exemplify this by 
choosing  parameter dependent values
$\varepsilon_i^{(1)}$ in the first channel and 
parameter independent values  $\varepsilon_i^{(2)}$ in the second
channel. In the following, we show a few  characteristic
results.

\subsubsection{Merge of states with gain and loss; second channel with two states}
\label{num2a}

We start these calculations with two channels by choosing 
two merged states with gain and loss  according to
Fig. \ref{fig2}. The second channel contains two resonance states
$i=1,2$ with negative widths and $|\gamma_i^{(2)}| \ll
|\gamma_i^{(1)}|$, see  Fig. \ref{fig5}.
 The eigenvalue and eigenfunction pictures 
 Fig. \ref{fig5}.a,b and c,d,e, respectively, 
 show the eigenvalues and eigenfunctions of the first channel 
 (Fig. \ref{fig2}.a-e) as well as the eigenvalues and eigenfunctions
 of the second channel. The last ones are constant as function of the
 parameter $a$ what follows from the assumption of their parameter
 independence. 
 
 The contour plots of the cross section
 (Figs. \ref{fig2}.f and \ref{fig5}.f) are different from one another.
 Common to both of them
 is that the cross section does not vanish in a finite range of the
 energy around $E \approx 0$ for all $a$.
 Under the influence of the two states in the second channel which are
 exactly in this energy and parameter range, the cross section is
 somewhat reduced. It does however not vanish.

These (and similar) simulations  show clearly the following
result. The fluctuations of the cross section which are caused by the
merging of two states with gain and  loss  in the first channel,
excite resonance states in the second channel. This happens although 
the nature of both channels is completely different.
In the first channel,  internal degrees of freedom of the system are
not excited,  while the
appearance of the resonance states in the second channel occurs via  
excitation of  internal degrees of freedom of the system.

\subsubsection{Level repulsion of states with gain and loss; second
  channel with one state}
\label{num2b}

In these calculations we start from the results shown in
Fig. \ref{fig4} for the first channel. The second channel contains
only one state. The eigenvalue and eigenfunction pictures
Fig. \ref{fig6}.a-e contain the eigenvalues and eigenfunction
trajectories of Fig. \ref{fig4}.a-e as well as those of the second
channel, e.g. the energy trajectories  at $E_i = 0.5$ and width
trajectories at $\Gamma_i /2 = -0.05$. The phase rigidity approaches
the value $1$ at $a=0$ in both cases.

The corresponding contour plot of the cross section is shown in
Fig. \ref{fig6}.f. It is related  to the contour plot of the first
channel   (Fig. \ref{fig4}.f);  and shows additionally the parameter
independent state of the second channel in the whole parameter range.

We performed further calculations with  parameters similar to those
used in   Fig. \ref{fig6}.a-e and show two of the corresponding
contour plots in  Figs. \ref{fig6}.g,h.
Both contour plots are obtained with
the comparably large width $\gamma_1^{(2)} /2 = - 0.5$;
the two energies $e_1^{(2)}$ are however different
from one another. The difference between  $e_1^{(2)}=0.5$
and  $e_1^{(2)}=-0.5$ can clearly seen in the corresponding contour plots
Figs. \ref{fig6}.g and h.

The results of these simulations with level repulsion in the first
channel and one state in the second channel
show the same characteristic features as those discussed above for
merging states. The fluctuations observed in the first channel are able
to excite resonance states in the second channel.

\subsubsection{Level repulsion of states with gain and loss; second
  channel with two states}
\label{num2c}

The situation with two states in the second channel is richer than
that with only one state because the two states can mix via the common
continuum.
 We show  results for one special case in Fig. \ref{fig7}. As in Figs.
 \ref{fig5}  and \ref{fig6}, the eigenvalue figures  \ref{fig7}.a,b
 contain the eigenvalue trajectories of both channels. The
 eigenfunction trajectories refer to the existence of an EP in the
 second channel: the phase rigidity of states related to the second 
channel is independent of the parameter $a$, as expected. It is 
however  smaller than $1$. Also the mixing $|b_{ij}|$ 
of the two wavefunctions via the common continuum 
in the second channel shows small deviations from the expectations. 
The relation of these results for the eigenfunctions of ${\cal
  H}^{(2,2)}$ to an EP in the second channel is discussed in detail in
appendix \ref{a1}. 
    
The contour plot  Fig. \ref{fig7}.f
shows the same characteristic features as  Fig. \ref{fig6}.f. Both,
its relation to the contour plot of the first channel
(Fig. \ref{fig4}.f)
as well to  the parameter independent state of the second channel  can
clearly be seen.
Thus, the fluctuations observed in the first channel  excite resonance
states in the second channel also in this case.

Summarizing the numerical results shown in the three figures 
\ref{fig5} to \ref{fig7} we state the following. 
Gain and loss in the first channel and excitation of resonance
states in the second channel are a  uniform process. This 
process can be described as a whole in 
the formalism for the description of open quantum systems which 
is used in the present paper.

\section{Discussion and summary of the results}
\label{disc}

In our paper, we considered gain and loss in an open quantum
system \cite{comment3}. Of special interest is the interplay
of these two opposed processes
in the neighborhood of singular points where it
causes fluctuations of the cross section. These fluctuations are 
observable and can excite resonance states in the system. 
The time scale of the fluctuations and that of the excitation of 
resonance states are very different.
The fluctuations of the cross section occur
quickly, without any excitation of internal degrees of freedom of the
system. The excitation
of resonance states is however much slower, and internal degrees of
freedom of the system are involved. 

The results of our calculations meet therefore the condition 
that the lifetime  of the primary process in
the  photosynthetic reaction center has to be very short.
Otherwise the energy  received from the photosynthetic
excitation, will change into heat and fluorescence \cite{lakhno}. 
This is a statement of Hermitian quantum physics, since the change of
energy into heat and fluorescence is impossible in 
non-Hermitian quantum
physics according to the results of our calculations. Here the
photosynthesis does not excite any eigenstate of the 
non-Hermitian Hamiltonian $\cal H$. Instead, the primay process occurs
due to fluctuations of the eigenfunctions of $\cal H$ around EPs.
The mechanism of photosynthetic excitation in non-Hermitian quantum
physics is therefore, as a matter of principle, different from that in
Hermitian quantum physics.  

We sketched first the formalism by means 
of which both the interplay
between gain and loss in an open quantum system \cite{comment3}
and the excitation of resonance states can be described 
as a uniform process. 
In any case, the widths $\gamma_i$ of the states are (generally) 
different from zero. In the first case we have two states with
different sign of the widths (corresponding to gain and loss),
while the states in the second case are standard decaying (resonance)
states with negative sign \cite{comment1}.
 
Generally, the cross section related to the two states with gain and
loss vanishes. Deviations from this rule appear
around the position of an eigenstate and in the 
neighborhood of singular (exceptional) points, where they may cause
non-vanishing fluctuations of the cross section.
These fluctuations have nothing in common with  resonances.
Rather, they are merely deviations from the vanishing value 
of the cross section and are  not related to the
excitation of any internal degrees of freedom of the system.  
They occur therefore with an  efficency of nearly 100 \%
at a very short time scale.

The fluctuations caused by the interplay of the states with 
gain and loss, are observable and may 
excite resonance states of the system after a comparably long time 
(which corresponds to the widths $\Gamma_i$ of these states).
The whole process of gain and loss together with the excitation of
resonance states is therefore characterized by 
two very different time scales: the quick process which creates the
fluctuations, and the slow process which is related
to the excitation of resonance states. The decay occurs thus
bi-exponential. Initially, it is determined by the quick process of
the interplay between gain and loss. Somewhere at its tail however, 
it will switch over to the slow process with excitation of internal 
degrees of freedom of the system.

The states with gain and loss as well as the resonance states excited
by the fluctuations, can each interact via a common environment
into which the corresponding states are embedded.
The two environments (channels)  will  never mix. 
This request is guaranteed in our formalism due to the very different 
time scales of the two processes. In any case,
this so-called  external mixing of the states occurs additionally
to the direct so-called  internal mixing of the states
which is supposed in our calculations to be involved in the 
complex energies
$\varepsilon_i \equiv e_i + i \gamma_i /2$  of the states.

According to the numerical results of our paper, the  fluctuations 
are   very robust. They appear in a relatively large
finite parameter range around the positions of the eigenstates 
and around EPs.

Finally, we mention a few interesting results which are characteristic
of open quantum systems including those considered in the present paper. 
\begin{enumerate}
\item[--]
The states of an open quantum system may interact via a common
environment into which the system is embedded. This mixing is called  
usually external interaction.  

\item[--]
The states of an open quantum system may have positive or negative
widths. The states with positive width \cite{comment1}
gain excitons (or information) from the
environment while those with negative width lose excitons 
(or information) due to
their coupling to the environment. As function of a parameter, gain
may   pass into loss and vice versa \cite{top}.

\item[--]
The phases of the eigenfunctions of a non-Hermitian operator are,
generally, not rigid, see Figs. \ref{fig1} to \ref{fig7} and 
\cite{top,proj10}.

\item[--]
Irreversible processes  determine the evolution 
of an open quantum system up to the occurrence of orthogonal eigenstates 
at maximum  width or level repulsion,  
see Figs. \ref{fig4}, \ref{fig6},  \ref{fig7} and \cite{top,proj10}.  

 \end{enumerate} 

We mention further that results similar to those discussed above 
for systems with two states, 
appear also in calculations for systems with more than two states, 
see \cite{proj10,pra93,epj2}. This holds true also for systems with
gain and loss.

\section{Conclusions}
\label{concl}

In our paper, we provided some results for a two-step
process which we obtained in the framework of the non-Hermitian
formalism  \cite{top,proj10} for the description of open
quantum systems. The first step is the interplay between
gain and loss of information  (excitons)
from an environment, while the second step is the 
excitation of a resonance state. The two steps are 
treated as two parts of the whole process. 

The total process might simulate photosynthesis:  the
first step is the capture of light in the light-harvesting complex
while the second step is the transfer of the excitation energy to the
reaction center which stores the energy from the photon in chemical
bonds.  That means: gain simulates the acceptor for light,
and loss stands for the donor which excites a
resonance state and simulates the coupling to the  sink. 
Altogether, the energy of the light is transferred  
to the reaction center of the light-harvesting complex.
The obtained results are very robust, and fluctuations play an 
important role.

The  results show some characteristic features 
which correspond, indeed, to those discussed 
in the literature for the photosynthesis.
Most interesting are the following results of our calculations:
\begin{enumerate}
\item 
 the  efficency of energy transfer is  nearly 100 \%;
 \item 
 the energy transfer takes place at a very short time scale;
 \item
  the storage of the energy in the reaction center occurs at a
much longer time scale;
\item 
according to points 2 and 3, the decay is bi-exponential.
\end{enumerate}

In future studies, the theoretical results have to be confirmed 
by application of the formalism to the description of concrete 
systems  in close cooperation between theory and experiment.

\vspace{.7cm}

{\bf Acknowledgment}

We are indebted to J.P. Bird for valuable discussions.

\vspace{.5cm}

\appendix

\section{Exceptional point in the second channel}
\label{a1}

At an EP the two eigenfunctions of a non-Hermitian Hamilton operator
$\cal H$ are exchanged
according to (\ref{sec8}). In more detail: tracing the eigenfunctions
of $\cal H$ as function of a certain parameter $b$, the
two eigenfunctions jump according to  (\ref{sec8})
at the critical parameter value $b=b^{\rm cr}$ which defines the
position of the EP. Some years ago, it has been shown 
\cite{ro01,ro03} that the influence of the EP is not  restricted 
to the jump occurring at $b^{\rm cr}$. It appears rather
in a  finite parameter range of $b$ in which the wavefunctions
of the two states are mixed according to
\begin{eqnarray}
\label{app1}
\Phi_i^{\rm ch} = \beta_k \Phi_k \pm i \beta_l \Phi_l 
\end{eqnarray}  
with $k \ne l$. 
The two wavefunctions $\Phi_i^{\rm ch}$ vary smoothly (i.e. without
any jump of the sign of their components) 
everywhere but at  $b=b^{\rm cr}$. 
Using the representation 
\begin{eqnarray}
\label{app2}
\Phi_i^{\rm ch} =|\Phi_i^{\rm ch}|e^{i\theta_i} 
\end{eqnarray} 
$\theta_i$ depends on the parameter $b$. After removing a common phase
factor, it follows $\theta_i^{(1)} \to \pi /4$ and $\pm 3 \pi /4$,
respectively, in approaching 
$b=b^{\rm cr}$; and $\theta_i^{(2)} \to 0$ or $\pi$ when $b$ is far
from the critical region around $b^{\rm cr}$.
In between the values  $\theta_i^{(1)}$ and $\theta_i^{(2)}$, 
the angle $\theta_i$  varies smoothly. 

Corresponding to the dependence of
$\theta_i$ on the parameter $b$, also the phase rigidity (\ref{eif7})
depends on this parameter in a certain finite
 parameter range. The phase rigidity is determined by the ratio
\begin{eqnarray}
\label{app3}
r_i \equiv
 \frac{{\rm Re}(\beta_k+\beta_l)^2}{|\beta_k + \beta_l|^2}
\end{eqnarray} 
which approaches $0$ at the EP (due to $|\beta_k + \beta_l|^2
\to \infty$) and  $1$ far from the EP (because here the wavefunctions 
are almost real).
The intermediate values of the phase rigidity $r_i$ 
are determined by the expression (\ref{app3}) 
calculated with the actual values $\beta_k$ and $\beta_l$. 
The results of these calculation give values for $r_i$ 
that are between the two limiting values $0$ and $1$.

Fig. \ref{fig7} shows
numerical results for the eigenfunctions of $\cal H$ 
which are obtained for calculations with two channels and 
two states in the second channel. We see
the values  $r_i$  for the states of the first channel 
(see Figs. \ref{fig2} and \ref{fig4})
as well as those for the two states of the second channel. 
They are independent of one another. The $r_i$ 
related to  the second channel are 
constant in the whole parameter range $a$  
shown in the figures (which is defined for the first channel).   
They may be smaller than 1, see  Fig. \ref{fig7}. 
  
The results obtained for the phase rigidity $ r_i$ of the two states
in the second channel,
may be considered as a proof of the finite parameter range
in which  an EP influences 
the properties of the system. The value  $ r_i$ is nothing but
an expression for  the distance of the system 
from an EP in the second channel: the larger $r_i$, the more distant
is the EP, while $r_i \to 0$ indicates that the EP is approached.
Thus, the  value $r_i$  in the eigenfunction pictures of
the two-channel system with two states in the second channel
additionally to those of the one channel system (Figs. \ref{fig2}  and
\ref{fig4}), 
allows us to determine the position of the EP in the second channel.


\begin{thebibliography}{99}
\bibitem{comment1} In contrast to the definition that is used in, 
for example, nuclear physics, we define the complex energies 
before and after diagonalization of ${\cal H}$ by 
$\varepsilon_k = e_k + \frac{i}{2} \gamma_k$ and  
${\cal E}_k = E_k + \frac{i}{2} \Gamma_k$, respectively, 
with $\gamma_k \le 0$ and $\Gamma_k \le 0$ for decaying states. 
This definition is useful when discussing systems with  gain
(positive widths) and  loss (negative widths).
\bibitem{comment2} The coalescence of two eigenvalues of a
  non-Hermitian operator should not be confused with the degeneration
  of two eigenstates of a Hermitian operator. The eigenfunctions of
  two degenerate states are different and orthogonal while those of
  two coalescing states are biorthogonal and differ only by a phase,
  see Eq. (\ref{sec8}).
\bibitem{comment5} Numerical calculations with $\gamma_i^{(2)}
\approx \gamma_i^{(1)}$ have shown that, in such a case, the states of
the second channel mix also via the first channel.
The obtained results are abstruse from the point of view of physics,
although they are mathematically correct.
\end{thebibliography}
\end{document}